\documentstyle[epsfig]{elsart}

\begin{document}
\begin{frontmatter}

%\begin{center}
\title{Event-by-event 
study of DCC-like fluctuation in ultra-relativistic nuclear collisions} 

\author{  M.M. Aggarwal$^{a}$, G. Sood$^{a}$ and 
Y.P. Viyogi$^{b}$ }\\
\small{a) Department of Physics, Panjab University , Chandigarh 160014, India\\
b) Variable Energy Cyclotron Centre, Bidhan Nagar, Kolkata 700064, India \\}

%\end{center}
\begin{abstract}

      A method based on  sliding window scheme
 is developed to search for patches in the pseudorapidity-azimuth plane,
on an event-by-event basis,
having unusual fluctuation in the
neutral pion fraction  which may arise due to the formation of Disoriented
Chiral Condensates (DCC) in high energy nuclear collisions. 
The efficiency of the method to extract the patches
and the purity of the extracted 
sample are studied for possible experimental situations.
%using a simple DCC model.
% The sensitivity of the method is found to be 
% limited only by the available statistics and can be several 
%order of magnitude better than those employed in earlier analyses.
% and at least an

\end{abstract}

\end{frontmatter}

Disoriented Chiral Condensates (DCC) have been predicted to be formed in high
energy hadronic and nuclear collisions when the chiral symmetry is 
temporarily restored at high temperatures. As the matter cools and expands,
the vacuum may relax into a state that has an orientation different from the
normal vacuum. This may lead to the formation of localized domains of DCC,
emitting  low momentum pions in a single direction in isospin 
space~\cite{review}. 
This would lead to event-by-event
fluctuation in the number of charged particles and photons in a given
phase space\footnote{A section of the pseudorapidity($\eta$)--azimuth($\phi$) 
phase space where DCC pions
may be localized will be called a patch.}, since majority of photons 
originate from $\pi^0$ decay and
charged particles are mostly charged pions.
It has been estimated that the 
neutral pion fraction($f$), corresponding to the DCC domain, follows the
probability distribution
$P(f) = \frac{1}{2\sqrt{f}}, {\rm where} 
  f = \frac{N_{\pi^0}}{N_{\pi^0}+N_{\pi^{\pm}}}.$
This distribution is markedly different from the standard
binomial distribution (peaking at $f$= $\frac{1}{3}$) for generic
pion production in nuclear collisions. 
The observed Centauro($N_{ch}$ $>$ $N_{\gamma}$)
and Anti-Centauro($N_{ch}$ $<$ $N_{\gamma}$)
events in cosmic ray collisions~\cite{jacee}
 may be due to the formation of large
domains of DCC.

There is very little theoretical guidance about the nature and
probability of DCC formation in nucleon-nucleon and nucleus-nucleus 
collisions. Crude theoretical estimates suggest that in central lead-lead
collisions at the CERN SPS, DCC formation probability may be
 ${}^<_\sim$10$^{-3}$~\cite{serreau}. The results of DCC search in
cosmic ray~\cite{pamir} and 
$\overline p p$ experiments~\cite{minimax} have been largely inconclusive.
In nucleus-nucleus collisions,
upper limits at 90\% confidence level have been set in the
analysis of data from the WA98 and NA49
experiments at the 
SPS~\cite{wa98global,wa98local,wa98central,na49}
 within the context of a  simple DCC model and under various assumptions.
For domains localized in ($\eta,\phi$) the most strict  limit set so far,
 for the top 5\% central Pb-Pb collision events
in the WA98 experiment~\cite{wa98central},
 are $\sim$10$^{-2}$ and $\sim$3$\times$10$^{-3}$ for
azimuthal domain sizes 45$^\circ$-90$^\circ$ and 90$^\circ$-135$^\circ$
respectively.

 Several methods have been proposed to detect the DCC formation in 
high energy hadron and heavy ion collisions and also used in the analysis
of experimental data. 
For details see the recent review
by Mohanty and  Serreau~\cite{review}.
The sensitivity of the methods has been studied using a simple DCC model and 
generating simulated data containing
a mixture of generic and DCC type events.
The Discrete Wavelet Transform  method~\cite{nandi}
is found to be sensitive if the fraction of DCC events is at
least 2$\times$10$^{-3}$ for domain size in the range of 40$^\circ$-90$^\circ$
 in a scenario where all the pions within the domain are of 
DCC origin. Considering only pions with $p_{\rm T} <$300~MeV/c
being of DCC origin, the sensitivity is found to be at 10$^{-2}$ level.
The fluctuation measure~\cite{mohantyD} has similar sensitivity.
$\Phi$-measure~\cite{mohantyphi}
is useful if each event contains 25\% pions from DCC source and the DCC event
fraction in the sample is 10$^{-2}$. 
It is thus found that the present methods, including those used in the
analysis of data from the WA98 experiment,
have sensitivity in the same range as the possible DCC formation probability. 
In this letter we describe a method which allows direct observation of
DCC-like patches by examining the event structure
 and its sensitivity can be better than 10$^{-4}$, being
 limited only by the available statistics.

We use a simple DCC model where
charged neutral fluctuation is introduced in generic events 
produced by the VENUS event generator~\cite{venus}, keeping initially
the identity of $\pi^0$ intact. It is 
assumed that the DCC pions survive till freezout, and that they also remain
localized in a ($\eta,\phi$) patch. To introduce DCC-like fluctuation
the identity of charged and neutral pions is interchanged pairwise
($\pi^{+}$$\pi^{-} \leftrightarrow  \pi^0$$\pi^0$) according to the 
$\frac{1}{2\sqrt f}$
probability distribution
 within a randomly selected ($\eta,\phi$) domain. 
 Two scenarios are considered for DCC-like fluctuation in the selected domain:
\begin{itemize}
\item  DCC-I : all the pions   are of DCC origin; 
\item DCC-II : only the pions having $p_{\rm T}\le $250 MeV/c
are of DCC origin. 
\end{itemize}

For the present 
study the DCC domain size is kept as one unit in pseudorapidity and 
60$^\circ$ in azimuth.
$\pi^0$'s are allowed to decay after introducing the DCC-like fluctuation.
Considering experimental situations where one 
measures the multiplicities of 
charged particles ($N_{ch}$) and photons ($N_{\gamma}$) in an event,
the neutral pion fraction 
may be approximately written as
$f =  \frac{N_{\gamma}/2}{N_{\gamma}/2+N_{ch}} $.
We consider  typical nuclear systems and  collision centralities such that
after $\pi^0$-decay, the mean multiplicities
of photons  and charged particles are $\sim$400  within one unit of
pseudorapidity and with full azimuthal coverage in the generic event. Such
multiplicities are typical for detector systems available in
 SPS and RHIC experiments.

\begin{figure}
\centerline{\includegraphics[scale=0.5]{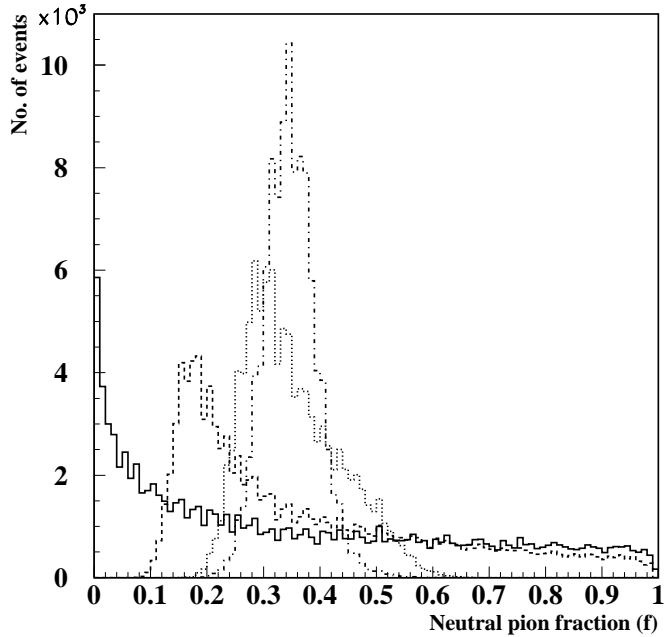}}
\caption{ Neutral pion fraction($f$) distributions for DCC-I patches 
before $\pi^0$-decay 
(solid line). The other three distributions are, after $\pi^0$-decay, 
for generic events (dash-dot line), DCC-I patches (dashed line) 
 and DCC-II patches 
(dotted line).}
\label{fig1}
\end{figure}

Neutral pion fraction ($f$) distribution for DCC-I patches 
before $\pi^0$-decay is shown in Fig.~\ref{fig1}. The
$f$-distribution after $\pi^0$-decay,
is also shown in Fig.~\ref{fig1} for various cases. In general
the $f$-distributions shrink inward for both scenarios of DCC formation
 after $\pi^0$ decay~\cite{poslim},
the effect being more pronounced for DCC-II.
It is clear that in the central part of $f$-distribution
the study of DCC-like fluctuation is difficult due to overwhelming background
from the binomial distribution of generic particle production mechanism.
We therefore concentrate on the low-$f$ and high-$f$ regions, which will be
referred to as charge-excess and photon-excess type fluctuations respectively.

Different ensembles of simulated DCC-like events are generated by adding 
events having
DCC-like fluctuation   and generic
events in different proportions, reflecting the DCC formation
probability. Such ensembles of events will be referred 
to as 'simulated data'. The fraction of DCC-like 
events ($\alpha$) in various ensembles of simulated
data varies from 2$\times$10$^{-4}$ to 1. 
The total number of events in each ensemble is 10$^5$.

The search for any non-statistical fluctuation requires a reference
 for comparison which can 
describe statistical fluctuation in a model-independent manner
and is free from inherent correlations.
For the analysis of experimental data the technique of 'mixed events' provides
such a reference set where correlations among 
particles are completely destroyed. The
mixed events are generated 
 using the standard procedure of taking large number 
of events (at least equal to the multiplicity of particles) and constructing
new events having particles picked up randomly, one particle from one event.
These events are created with the same multiplicity distribution as the 
'simulated data', similar statistics
 and have similar charge-neutral correlation on a global scale.

For the study of charged neutral fluctuation, we assume that there are
detectors measuring the photon and charged particle multiplicities in
overlapping part of ($\eta,\phi$) phase space, with or 
without the $p_{\rm T}$-information of particles. 
The common coverage of the two detectors is
assumed to be one unit in $\eta$ and 2$\pi$ in $\phi$,
the introduced DCC domain being fully contained within this region.
We develop a simple but powerful 'sliding window' method (SWM) in which a 
window in azimuthal plane, say of size $\Delta\phi$, is chosen in the common 
coverage of the two detectors
in which the neutral pion
fraction $f$ is calculated. The entire azimuthal
range of common coverage 
is scanned by continuously sliding the window, shifting each time by a 
small amount, say $\delta\phi$, to search for a patch having a neutral pion
fraction several standard deviations away from the mean value. This method
utilizes the full advantage of azimuthal resolution of the detectors and allows
direct observation of patches having large (or small) $f$-values. The value of
$\delta\phi$ depends on the azimuthal resolution of the two detectors. For the
present work, we have chosen $\delta\phi$=2$^\circ$. A preliminary version of 
the SWM is described in \cite{mmapramana}.

\begin{figure}
\centerline{\includegraphics[scale=0.5]{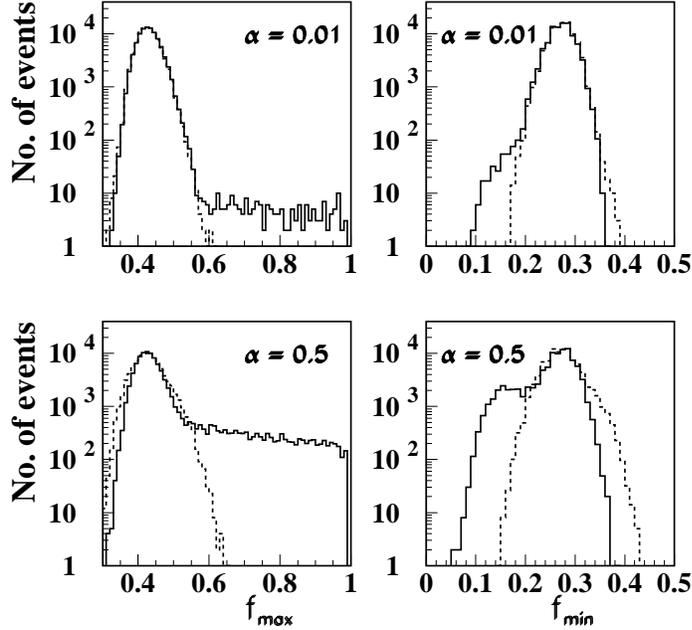}}
\caption{ $f_{\rm max}$(left) and $f_{\rm min}$(right) distributions  
for a $\Delta\phi$ =
 $60^{\circ}$ window size for two values of $\alpha$ in DCC-I scenario
as indicated. Continuous lines : simulated events, dashed lines : mixed events
generated from corresponding simulated events.}
\label{fmaxfmin}
\end{figure}

The SWM provides a set of $f$-values in each event. We focus on the
maximum and minimum $f$-values ($f_{\rm max}$ and $f_{\rm min}$)  
to search for photon-excess and 
charge-excess type fluctuations. The $f_{\rm max}$ and $f_{\rm min}$ 
distribution of simulated data for $\alpha$=0.01 and 0.5 
are shown in Fig.~\ref{fmaxfmin} for a
typical window size $\Delta\phi$=60$^\circ$ along with those for
mixed events generated from those samples. The
$f_{\rm max}$ and $f_{\rm min}$ distributions for mixed events are 
almost Gaussian in shape for lower value of $\alpha$ but
become skewed for  large values of
$\alpha$.

It is observed that the signals of non-statistical fluctuation can be 
distinguished only in the outer regions. We estimate the mean ($\mu$)
 and r.m.s. deviation ($\sigma$) of the
mixed event distributions. A cut on the $f_{\rm max}$ and $f_{\rm min}$ 
distribution of simulated data
is applied by $\mu \pm n\sigma$, where $n$ ranges from 3 to 5, 
positive sign
being applied for $f_{\rm max}$ and negative sign for $f_{\rm min}$
distributions. The
events having patches where the $f_{\rm max}$ or $f_{\rm min}$ values are
 beyond the cut are labeled as 'extracted'. 

%\section{The results}

%\subsection{Statistical fluctuation}

The sensitivity of the SWM to extract DCC-like fluctuation is decided by 
the limit of statistical background that the mixed event sample gives for any 
given extraction procedure. 
The patches in mixed events having very large (or very small) $f$-values, 
arising 
due to statistical fluctuation, constitute the background to the study of
DCC-like fluctuation. The statistical fluctuation depends on the multiplicity
of particles in the patches and hence also on the window size in the SWM.

\begin{figure}
\centerline{\includegraphics[scale=0.5]{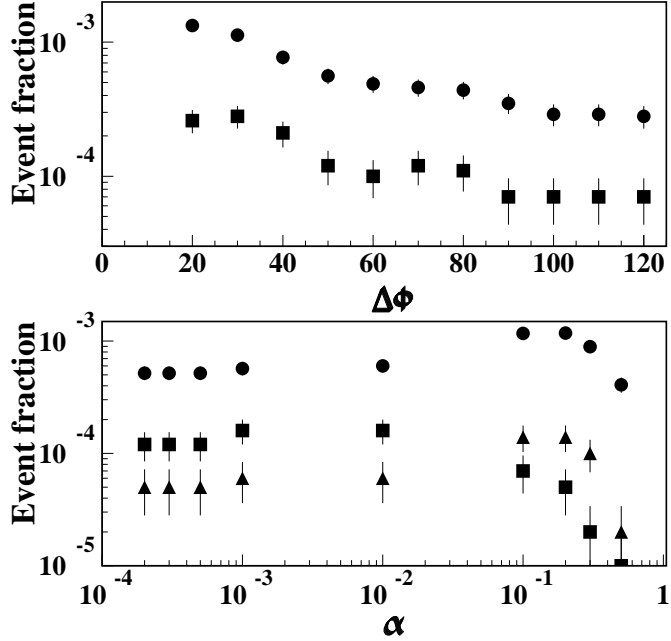}}
\caption{Fraction of events  versus window size for $\alpha$=0 (top) and 
 versus $\alpha$ for $\Delta\phi$=60$^\circ$ (bottom) 
for 
$f_{\rm max}$ distribution with $\mu$+4$\sigma$ cut (filled circles)
 and for $f_{\rm min}$ distribution with
$\mu$-4$\sigma$ cut (filled squares)
for mixed events generated
from different  ensembles of events. 
Filled triangles in the bottom panel represent fraction of events for
$f_{\rm max}$ distribution with
$\mu$+5$\sigma$ cut.}
\label{statfluct}
\end{figure}

For mixed events produced from generic events ($\alpha$=0) 
Fig.~\ref{statfluct}(a)
 displays  the fraction of events, which constitutes the background, 
versus window size for $\mu\pm 4\sigma$ cut on mixed event $f_{\rm max}$ and 
$f_{\rm min}$ distributions. 
The background is almost independent of window size for
$\Delta\phi {}^>_\sim$ 60$^\circ$ and is found to be 
4$\times$10$^{-4}$
 and 1$\times$10$^{-4}$
 respectively for photon-excess and charge-excess cases.
For subsequent investigations we have used $\Delta\phi$=60$^\circ$.

The background due to statistical fluctuation has also been investigated 
using the mixed 
events generated from different ensembles of events (different $\alpha$ 
cases), 
keeping the window size fixed at 60$^\circ$. Fig.~\ref{statfluct}(b)
 shows the fraction of 
events  as a function of $\alpha$ for $\mu \pm 4\sigma$ cuts on mixed event
$f_{\rm max}$ and $f_{\rm min}$ distributions. The 
background  is of similar magnitude as that in Fig.~\ref{statfluct}(a) for 
lower values of $\alpha$. With increasing $\alpha$, the background increases 
and then decreases for very high values. This apparent decrease at very high 
$\alpha$ is due to the distorted shape of the corresponding mixed event 
distributions and the resulting high values of $\sigma$. 
 Fig.~\ref{statfluct}(b) also displays the background for 
photon-excess cases with 
$\mu +5\sigma$ cuts.
The background values are now an order of magnitude smaller, 
being as low as 5$\times$10$^{-5}$ for smaller values of $\alpha$.

\begin{figure}
\centerline{\includegraphics[scale=0.5]{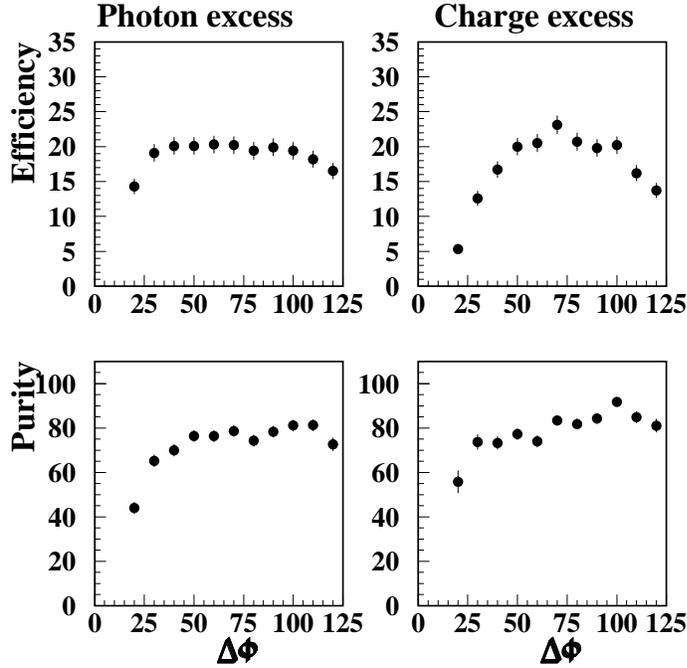}}
\caption{  Efficiency and 
purity (expressed in percentage) 
versus window size  for simulated data with $\alpha$=0.01 and
$\mu\pm 4\sigma$ cut as the extraction criteria. Left panels are for
 photon-excess cases and right panels for charge-excess cases.}
\label{effpurwin}
\end{figure}

In the present case of simulation study,
where the character of each event and patch is known,  we can estimate the
'background' (B) which arises from the following sources : (a) the
 extracted patch
 is from a generic event and (b) the extracted patch, in the DCC-like event,
is not from the azimuthal region where DCC was actually introduced. 
Extracted patches which belong to the DCC-like events and are found around the
same azimuthal location are labeled as 'extracted signal' (S).
In order to characterize the SWM
we estimate the efficiency of extraction S/N, where N is the number
of DCC-like events present in the ensemble depending on $\alpha$ value, and 
 the 'signal purity of extracted patches' S/(S+B).

The effect of window size on the results of SWM analysis has been investigated 
further for the case of simulated events with $\alpha$=0.01. The efficiency of 
extraction and the purity of the extracted patches are plotted for the case of 
both charge-excess and photon-excess fluctuations as a function of window size 
in Fig.~\ref{effpurwin}, using $\mu \pm 4\sigma$ cut 
as the extraction criteria. It is 
found that the efficiency and purity are not much affected by varying the 
window size over a wide range around the azimuthal domain where fluctuation 
was originally introduced. The results are affected only at very small window 
sizes, which, as discussed earlier, get dominated by statistical fluctuation 
due to small number of particles.

\begin{figure}
\centerline{\includegraphics[scale=0.5]{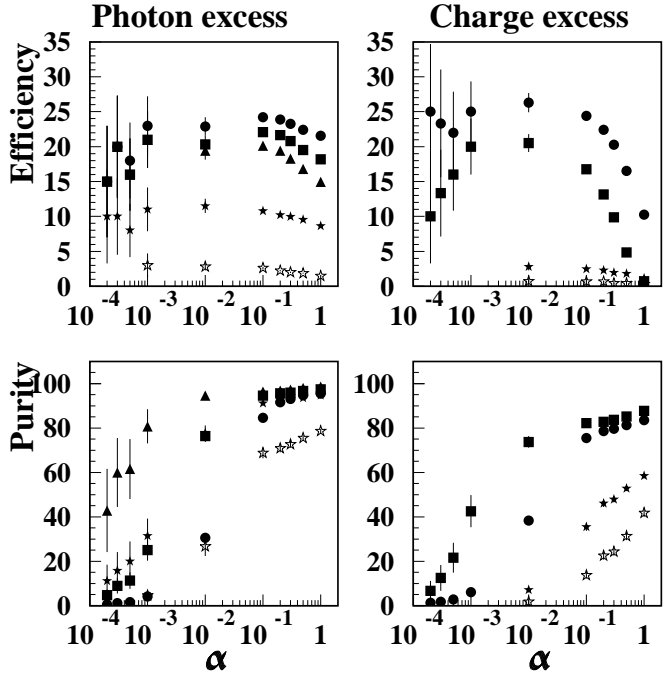}}
\caption{Efficiency and purity versus $\alpha$. The left panels are for
photon-excess and the right panels for charge-excess fluctuations. Filled
circles : 3$\sigma$ cut for DCC-I, filled square : 4$\sigma$ cut for DCC-I,
filled triangles : 5$\sigma$ cut for DCC-I, 
open stars : 4$\sigma$ cut for DCC-II with
multiplicity measurement only, filled stars : 4$\sigma$ cut for DCC-II
with the addition of charged particle $p_{\rm T}$ measurement.}
\label{effpurall}
\end{figure}

Fig.~\ref{effpurall} shows the efficiency and purity as a function of 
$\alpha$ for the entire data set analyzed in the present work for both 
charge-excess and photon-excess type fluctuations in DCC-I 
and DCC-II scenarios. The extraction criteria
applied is a cut of $\mu\pm n\sigma$ on the $f_{\rm max}$ and 
$f_{\rm min}$ distributions obtained from the corresponding mixed events for
each $\alpha$. Results for DCC-I scenario 
are presented for $n$=3,4,5 for photon-excess and for
$n$=3,4 for charge-excess cases.

As a function of increasing $n$, the signal falls very slowly whereas the 
background falls sharply. The efficiency is almost constant for lower values
of $\alpha$. The decrease in efficiency for larger values of $\alpha$ is due to
distortions in the mixed event $f_{\rm min}$ and $f_{\rm max}$
distributions and the apparently large $\sigma$-values.
The  efficiency for extraction of patches for both 
the charge-excess and photon-excess cases in DCC-I scenario
 is found to be in the range of
 10-20\%, lower values of efficiency corresponds to higher purity.   
Even for $\alpha$ as low as 3$\times$10$^{-4}$, the purity
is more than 50\%, indicating that the background is less than the signal. It
is clear that if the statistics is high, one can make even tighter cuts 
($n >$5) to further improve the purity with only marginal loss in
 efficiency.

For the analysis of data having fluctuation in DCC-II scenario, two approaches 
are used. First we assume that only multiplicity information is available. The 
analysis proceeds in a manner identical to what was done for DCC-I case. 
The efficiency of extracting DCC-II type patches and the purity of such 
extracted patches are
 shown in Fig.~\ref{effpurall}
using $\mu \pm 4\sigma$ cut as the extraction criteria. 
The efficiency is now
 quite low for both charge-excess and photon-excess type patches. This is 
understandable as the fraction of DCC pions in the patches is now reduced to 
less than half.

In the second approach we assume that the $p_{\rm T}$ information of charged 
particles is available. Such a possibility exists in the STAR experiment at 
RHIC~\cite{ftpcpmd}. The neutral pion fraction $f$ is now calculated by taking
 only the charged particles having $p_{\rm T} <$250~MeV/c. 
Using the new $f$-values
both $f_{\rm min}$ and $f_{\rm max}$ distributions are generated.
The efficiency 
of extraction and the purity of extracted samples, shown in 
Fig.~\ref{effpurall}
using the $\mu \pm 4\sigma$ cut as the extraction criteria, 
are found to increase by a factor of 
about 5-10 over the case when only multiplicity information was used.

In summary the SWM provides a simple and elegant method for the study of
charge-neutral fluctuation in ultra-relativistic nuclear
 collisions. The sensitivity of the
method, decided by the background due to statistical fluctuation, can be 
improved to a great extent by making suitable extraction criteria so that the
background is minimized. In the present analysis, with just 10$^5$ events,
it has been possible to reduce the background to 5$\times$10$^{-5}$.
 This is almost
 two orders of magnitude better than the previous methods. At higher 
multiplicity, as is expected at the LHC experiments, background 
due to statistical fluctuation may be further
reduced. 

The efficiency of
extracting the DCC-like patches varies from  a few percent to about 40\% 
(combining the photon-excess and charge-excess cases). This  
depends  on the extraction criteria,
 DCC formation probability and the $p_{\rm T}$ distribution of
DCC pions. The purity of the extracted patches also depends  on those
parameters and can reach 60-80\% even for low values of $\alpha$. For high
values of $\alpha$ this reaches close to 100\%.
For a given $\alpha$ and extraction criteria the purity is higher for photon
excess cases than for charge-excess cases as observed
earlier~\cite{poslim}. In DCC-II scenario the efficiency
is much lower for charge-excess cases than for
photon-excess cases. Considering limitations due to detector 
effects~\cite{poslim} it is advantageous to look for photon-excess type
charge-neutral fluctuation.

The SWM is a general method and can be utilized not only in azimuthal space
but also in pseudorapidity space or even in any combined phase space with
multi-dimensional windows and using any suitable
physical observable which can be computed over the window.
The only important point is to slide the window over 
the acceptable phase space region to compute a number of values of the
observable under investigation and then study their distributions. This allows
for direct observation of unusual structures in the event
and draw conclusions which are 
model-independent.

The method developed here can be directly applied to the analysis of experimental data using mixed event as the reference set  without recourse to
any DCC model.
With improved statistics, which is now-a-days available in collider 
experiments, it should be possible to reduce the background to any desired 
level  and improve the sensitivity of the SWM. This should also allow one to
employ tight cuts which further improves the signal purity  to the
estimated level to confirm or deny the occurrence of exotic phenomena.

 The financial assistances from the Department of Atomic Energy
 and Department of Science and Technology of the Government of India
are gratefully acknowledged.

\end{document}